\documentclass[journel]{IEEEtran}
\IEEEoverridecommandlockouts
\usepackage{cite}
\usepackage{amsmath,amssymb,amsfonts}
\usepackage{mathrsfs}
\usepackage{graphicx}
\usepackage{textcomp}
\usepackage{color}
\usepackage{epstopdf}

\def\BibTeX{{\rm B\kern-.05em{\sc i\kern-.025em b}\kern-.08em
    T\kern-.1667em\lower.7ex\hbox{E}\kern-.125emX}}

\begin{document}

\title{$\mathscr{H}_2$ Model Reduction for Linear Quantum  Systems\\
}
\author{\IEEEauthorblockN{ Guangpu Wu},
\IEEEauthorblockN{ Shibei  Xue},
\IEEEauthorblockN{ Guofeng Zhang}, and
\IEEEauthorblockN{Ian R. Petersen}

\thanks{``This work is supported by the National Natural Science Foundation of China (NSFC) under Grants No. 62273226 and No. 61873162.'' }
\thanks{Guangpu Wu is with the Department of Automation, Shanghai Jiao Tong University, Shanghai 200240, P. R. China (email:\emph{18716032768@sjtu.edu.cn}) }
\thanks{Shibei Xue is with the Department of Automation, Shanghai Jiao Tong University, Shanghai 200240, P. R. China (email:\emph{shbxue@sjtu.edu.cn}). }
\thanks{Guofeng Zhang is with the Department of Applied Mathematics, The Hong Kong Polytechnic University, Hung Hom, Kowloon, Hong Kong Special Administrative Region, China (email:\emph{guofeng.zhang@polyu.edu.hk}).}
\thanks{Ian R. Petersen is with the School of Engineering,  Australian National University, Canberra, ACT 2601, Australia (email:\emph{i.r.petersen@gmail.com}).}
}
%
%

\maketitle

\begin{abstract}
In this paper, an $\mathscr{H}_2$ norm-based model reduction method for linear quantum systems is presented, which can obtain a physically realizable model with a reduced order for closely approximating the original system. The model reduction problem is described as an optimization problem, whose objective is taken as an $\mathscr{H}_2$ norm of the difference between the transfer function of the original system and that of the reduced one.
Different from classical model reduction problems, physical realizability conditions for guaranteeing that the reduced-order system is also a quantum system should be taken as nonlinear constraints in the optimization.
  To solve the optimization problem with such nonlinear constraints,
  we employ a matrix inequality approach to transform nonlinear inequality constraints into readily solvable linear matrix inequalities (LMIs) and nonlinear equality constraints,
 so that the optimization problem can be solved by a
 lifting variables approach.
We emphasize that different from existing work, which only introduces a criterion to evaluate the performance after model reduction, we guide our method to obtain an optimal reduced model with respect to the $\mathscr{H}_2$ norm.
In addition, the above approach for model reduction is extended to passive linear quantum systems.
Finally, examples of active and passive linear quantum systems validate the efficacy of the proposed method.
\end{abstract}


\section{Introduction}
Linear  quantum  systems play a crucial role in advancing applications such as quantum computing, quantum communication~\cite{3}, and quantum sensing~\cite{4}. Many applications involve several  linear quantum system components, which results in a high-dimensional total system.
 For example, linear quantum systems are widely used in applications such as cavity opto-mechanical systems \cite{4.01} and multi-mode quantum harmonic oscillators \cite{4.02}. These systems usually consist of multiple quantum states whose interactions form complex coupling networks \cite{4.03,4.04}.
Furthermore, in the augmented system for non-Markovian quantum systems\cite{4.05,17,13.1,13.2}, many linear quantum systems are used for modelling quantum colored noise, which leads to a high dimensional Hilbert space for the augmented system. Hence, it would be difficult to control these systems in real time since the high dimension of the system leads to a heavy computation burden for controllers\cite{5,5.1,5.2}.

%
In the above contexts, it is necessary to develop model reduction methods which can approximate a high-dimensional system with a lower-dimensional model as well as retain the essential dynamics of the original system \cite{14.9,10}, thus making the design of control algorithms more feasible and efficient. However, classical model reduction methods cannot be applied directly to linear quantum systems since
the matrices in their linear dynamical equations should satisfy physical realizability conditions~\cite{4.1,4.2}. These conditions resulting from commutation rules in quantum mechanics~\cite{5} guarantee the quantumness of a system, which is essential for outperforming its classical counterparts. Hence, model reduction methods for quantum systems should be developed.

In early research, model reduction methods were limited to singular perturbation techniques  for passive linear quantum systems \cite{11}, which can preserve  desired parts by separating fast changing parts from slow changing parts.
To develop a more general method, several effective classical techniques have been extended to linear quantum systems including  balanced truncation methods and interpolation projection methods~\cite{13,14}.
By identifying the energy balance state of a system, the balanced truncation method projects the system into a subspace with higher energy, thereby generating a simplified model that retains its essential dynamics.
In contrast to singular perturbation techniques, balanced truncation methods are capable of preserving stability and also provide error bounds, which serve to control approximation errors \cite{13}.
%
Additionally, similar to the interpolation projection methods used in classical model reduction, interpolation projection methods for linear quantum systems can ensure that the input-output responses of the original and reduced-order systems match at multiple selected frequencies.
Ref. \cite{14} introduces a tangential interpolation projection method for model reduction of linear quantum systems and establishes \(\mathscr{H}_\infty\) error bounds for the proposed method, incorporating a heuristic algorithm for the selection of tangential directions. However, this method only guarantees minimal error near the interpolation points and does not achieve accurate approximation across the entire frequency band.
 Hence, although numerous methods for model reduction exist, most of which can only evaluate the performance of methods after model reduction, they often fail to achieve optimum performance with respect to a specific metric.


In this paper, we consider $\mathscr{H}_2$ norm  as a criterion for model reduction.
Compared with existing model  reduction methods,
we not only use $\mathscr{H}_2$ norm as the criterion to evaluate model reduction, but also design the reduced model according to $\mathscr{H}_2$ performance.
Unlike in classical systems, obtaining an $\mathscr{H}_2$ optimal reduced model while ensuring physical realizability in linear quantum systems is a significant challenge. To address this issue, we first reformulate the model reduction problem as an optimization problem and rigorously derive the necessary conditions for a feasible solution.
 By utilizing the matrix inequality method, we transform the nonlinear constraints introduced by physical realisability into LMIs and nonlinear equality constraints, which allows us to solve the optimization problem using a matrix lifting method.
A method for model reduction of passive linear quantum systems is also presented.
Finally, the proposed method is applied to active and passive linear quantum systems, illustrating its effectiveness.

This paper is organized as follows. Section 2 introduces  linear quantum systems and the model reduction problem. Section 3 introduces the $\mathscr{H}_2$ model reduction method.
In Section 4, we present the model reduction approach for passive linear quantum systems.
Examples are given in Section 5 to demonstrate the effectiveness of our method. Finally, conclusions are drawn in Section 6.

%
%

\textbf{Notation} For a matrix $A=[A_{ij}]$, the symbols $A^T$, $A^\dag$ and $\text{tr}(A)$ represent the transpose, Hermitian conjugate and trace of $A$. We denote $\text{He}(A)=A+A^\dag$.  Given two operators $N_1$ and $N_2$, $[N_1,N_2]=N_1N_2-N_2N_1$ is their commutator. Given a complex number $a$, $\bar a$, $\mathfrak R\{a\}$ and $\mathfrak I\{a\}$ represent its conjugate, real part and  imaginary part, respectively. The signal $\#$ in the matrix represents symmetric elements.
\section{Problem formulation}\label{Problem}
Linear quantum systems are a cluster of interacting quantum harmonic oscillators which can be driven by bosonic quantum fields~\cite{5}.
 Both the modes of the system and the bosonic fields can be described by their corresponding canonical position and momentum operators, respectively. In the Heisenberg picture, these systems are characterized by linear dynamics governed by the canonical commutation relations of quantum mechanics \cite{4.04, 5.1}.

Concretely, we consider $x = (q_1, p_1, q_2, p_2, \ldots, q_n, p_n)^T$ as a vector of the canonical position and momentum operators of quantum harmonic oscillators. Here, $q_k$ and $p_k$ are position and momentum operators for the $k$-th oscillator satisfying  commutation relations $[q_k,p_k]=i\hbar$, where $\hbar$ is reduced Planck constant, such that the vector $x$ satisfies the canonical commutation relations (CCR) $xx^T - (xx^T)^T = i\mathbb{J}_n$~\cite{18}, with $\mathbb{J}_n = I_n \otimes \begin{bmatrix}0 & 1 \\ -1 & 0\end{bmatrix}$, where $I_n$ is an $n$-dimensional identity matrix.
In addition, we denote the incoming $m$ input continuous-mode bosonic quantum fields as $\mathcal{A}(t) = (\mathcal{A}_1(t), \ldots, \mathcal{A}_m(t))$, where $\mathcal{A}_j(t)$ are field operators associated with the bosonic input channels.
Generally, for a channel of input fields, there is a corresponding output field \cite{14.1}. So for $m$ channels of input fields, there are $m$ channels of output fields. However, in many applications, we are only concerned with part of the outputs. Hence,
we may consider $l\leq m$ channels of output fields which are denoted as
$\mathcal Y(t) = (\mathcal Y_1(t), \ldots, \mathcal Y_{l}(t))$.
 Here, $\mathcal{A}(t)$ and $\mathcal{Y}(t)$ represent quantum Wiener processes, which are fundamental in describing continuous-mode bosonic quantum fields.
 With these operators, the dynamical equation of a linear quantum system can be written as~\cite{11}:

\begin{align}\label{dx2}
&dx(t)=Ax(t)dt+Bdw(t),\nonumber\\
&dy(t)=Cx(t)dt+Ddw(t),
\end{align}
where $A\in \mathbb R^{2n\times2n}$, $B\in \mathbb R^{2n\times2m}$, $C\in \mathbb R^{2m\times2n}$, $D\in \mathbb R^{2l\times2m}$,
\begin{align}
w(t)=&(\mathfrak R\{\mathcal A_1(t)\},\mathfrak I\{\mathcal A_1(t)\},\mathfrak R\{\mathcal A_2(t)\},\nonumber\\
&\mathfrak I\{\mathcal A_2(t)\},...,\mathfrak R\{\mathcal A_m(t)\},\mathfrak I\{\mathcal A_m(t)\})^T,
\end{align}
and
\begin{align}
y(t)=&(\mathfrak R\{\mathcal Y_1(t)\},\mathfrak I\{\mathcal Y_1(t)\},\mathfrak R\{\mathcal Y_2(t)\},\nonumber\\
&\mathfrak I\{\mathcal Y_2(t)\},...,\mathfrak R\{\mathcal Y_{l}(t)\},\mathfrak I\{\mathcal Y_{l}(t)\})^T,
\end{align}
are the input  and  output vectors, respectively.  It is straightforward to obtain the input-output transfer function of the system as $\Xi_G(s)=C(sI_n-A)^{-1}B+D$, where $s$ is the complex variable in Laplace transform.

 Different from classical linear systems, the system \eqref{dx2} satisfies the following  physically realizability conditions~\cite{4}:
\begin{align}\label{real}
A\mathbb J_n+\mathbb J_nA^T+B\mathbb J_mB^T&=0,\\
\mathbb J_nC^T+B\mathbb J_m D^T&=0,\label{real2}\\
D\mathbb J_m D^T &=\mathbb J_{l},\label{real3}
\end{align}
such that the system \eqref{dx2} represents a quantum system.
We take the above system as the original system which is physically realizable and also Hurwitz stable.
 It should be noted that the system is $2n$-dimensional and this model is  capable of representing both passive and active linear quantum systems.
%

In practice, the system \eqref{dx2} would contain some redundant modes or the high dimension of the system \eqref{dx2} would lead to a large computational burden as mentioned above. Hence, it is necessary to explore model reduction techniques for the system \eqref{dx2}.
  When we consider the model reduction problem, we only consider reducing the number of modes of the system \eqref{dx2} so that the number of channels of input and output fields are kept the same as those of the original system \eqref{dx2}.
So we write the state vector of a $2r$-dimensional reduced system as $x_r(t)$ with  $r<n$ , which is in a similar form as that of $x(t)$  and the dynamical equation of the reduced model is described as
\begin{align}\label{dxr}
&dx_r(t)=A_rx_r(t)dt+B_rdw(t),\nonumber\\
&dy_r(t)=C_rx_r(t)dt+D_rdw(t),
 \end{align}
 where $A_r\in \mathbb R^{2r\times2r}$, $B_r\in \mathbb R^{2r\times2m}$, $C_r\in \mathbb R^{2m\times2r}$ and
  \begin{align}
 D_r=D. \label{DrD}
 \end{align}
 Likewise, we require that the reduced model \eqref{dxr}  be Hurwitz stable.
 Similarly, the transfer function of the reduced system can be given as $\Xi_{G_r}(s)=C_r(sI_r-A_r)^{-1}B_r+D_r$.
%
%

To enable the system (\ref{dxr}) to represent a quantum system, the system matrices should also satisfy the physical realizability conditions
  \begin{align}\label{realr}
A_r\mathbb J_r+\mathbb J_rA_r^T+B_r\mathbb J_mB_r^T&=0,\\
\mathbb J_rC_r^T+B_r\mathbb J_m D_r^T&=0,\label{realr2}\\
D_r\mathbb J_m D_r^T& =\mathbb J_l.\label{realr3}
\end{align}
Due to \eqref{DrD}, the physical realizability condition \eqref{realr3} is directly satisfied. In the following section, when we consider model reduction, we do not take \eqref{realr3} into account.

Different from existing work, we consider the $\mathscr{H}_2$ norm of the transfer function
\begin{align}\label{H2}
||\Xi_G(s)-\Xi_{G_r}(s)||_2 = \left( \frac{1}{2\pi} \int_{-\infty}^{\infty} \text{tr}\left[ \hat \Xi(j\omega)\hat \Xi^*(j\omega) \right] d\omega \right)^{\frac{1}{2}},
\end{align}
 with $\hat \Xi(s)=\Xi_G(s)-\Xi_{G_r}(s)$ as the objective, when we consider model reduction for retaining the key dynamics of the original system.
  It is emphasized that the $\mathscr{H}_2$ norm is not only used to evaluate the performance after model reduction, instead we also utilize it to guide the direction in the process of model reduction.

 Therefore, the model reduction problem considered in this paper is described as follows.

 Given the original stable system \eqref{dx2} satisfying the physical realizability conditions \eqref{real} and \eqref{real2}, the model reduction problem is to find a reduced order stable model described by the triple $A_r$, $B_r$, $C_r$ satisfying the physically realizability conditions \eqref{realr}, \eqref{realr2} and \eqref{realr3} such that the $\mathscr{H}_2$ norm $||\Xi_G-\Xi_{G_r}||_2$ is  minimized; i.e.,
 \begin{align}\label{newopt}
    & \min_{ A_r, B_r, C_r} J=||\Xi_G-\Xi_{G_r}||_2  \\
     & \text{s.t.} \quad
  \text{Eqs.}\  \eqref{realr},\ \eqref{realr2}.\nonumber
\end{align}


\section{$\mathscr{H}_2$ Model Reduction Method}
\subsection{Optimization Framework for Model Reduction}
\begin{figure}[htbp]
\centerline{\includegraphics[width=7.5cm]{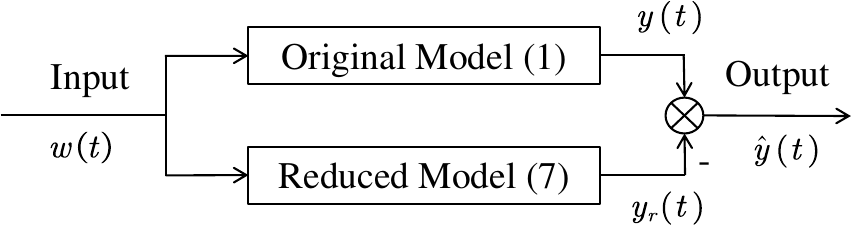}}
\caption{Block diagram of the augmentation system \eqref{dx3}}
\label{fig0}
\end{figure}

To solve the $\mathscr{H}_2$ model reduction problem \eqref{newopt}, we first provide necessary conditions in the optimization framework.
In order to evaluate the error between the original model \eqref{dx2} and the reduced model \eqref{dxr}, we construct an augmented system as shown in Fig. \ref{fig0}, which can be written as
\begin{align}
&d\hat x(t)=\hat A\hat x(t)dt+\hat Bdw(t),\label{dx3}\\
&d\hat y(t)=\hat C\hat x(t)dt,\label{dx4}
\end{align}
 where
\begin{align}
\hat x(t)&=\left[\begin{array}{c}x(t)  \\x_r(t)\\\end{array}\right],\nonumber\\
\hat A&=\left[\begin{array}{cc}A&0  \\0 &A_r\\\end{array}\right],\nonumber\\
\hat B&=\left[\begin{array}{c}B\\B_r\\
\end{array}\right],\nonumber\\
\hat C&=\left[\begin{array}{cc}C &-C_r\\\end{array}\right],\nonumber\\
\hat y(t)&=y(t)-y_r(t).\nonumber
\end{align}

It is not difficult to observe that the input term in Eq. \eqref{dx4} is eliminated due to Eq. \eqref{DrD}.

The stability of the augmented system \eqref{dx3} can be guaranteed since  both the system model \eqref{dx2} and the reduced model \eqref{dxr} are assumed to be Hurwitz stable.
Hence, we can find positive definite matrices $P$ and $Q$ that satisfy Lyapunov equations
\begin{align}
    &\hat A P + P \hat A^T + \hat B \hat B^T   = 0, \label{P} \\
    &\hat A^T Q + Q \hat A + \hat C^T \hat C  = 0. \label{Q}
\end{align}
For further derivation, we rewrite the matrices $P$ and $Q$  in a block form
\begin{align}
P=\left[\begin{array}{cc}P_1&P_2  \\P_2^T &P_3\\\end{array}\right],
Q=\left[\begin{array}{cc}Q_1&Q_2  \\Q_2^T &Q_3\\\end{array}\right],
\end{align}
 where $P_1, Q_1\in \mathbb R^{2n\times2n}$, $P_2, Q_2\in \mathbb R^{2r\times2n}$ and $P_3, Q_3\in \mathbb R^{2r\times2r}$.

In addition, the square of the $\mathscr{H}_2$ norm  \eqref{H2} can be calculated as \cite{15}
   \begin{align}
||\Xi_G-\Xi_{G_r}||^2_2={\rm tr}(\hat B^TQ\hat B)={\rm tr}(\hat CP\hat C^T).\label{XiG}
\end{align}
Therefore, we can transform the $\mathscr{H}_2$ model reduction problem into solving the following optimization problem.

For a stable linear quantum  system  \eqref{dx2} satisfying the physical realizability conditions \eqref{real}, \eqref{real2} and \eqref{real3},  find a Hurwitz matrix $A_r$, real-valued matrices $B_r$, $C_r$ and positive definite matrices $P$ and $Q$ that minimize the square of the $\mathscr{H}_2$ norm \eqref{XiG} satisfying  physical realizability conditions \eqref{realr} and \eqref{realr2} and Lyapunov equation constraints \eqref{P}, \eqref{Q}; i.e.,
\begin{align}
    & \min_{P, Q, A_r, B_r, C_r}J={\rm tr}(\hat B^TQ\hat B)={\rm tr}(\hat CP\hat C^T) \label{opt} \\
     &~~~~~~~~\text{s.t.}\
  \text{Eqs.}\   \eqref{realr}, \eqref{realr2},  \eqref{P}, \eqref{Q}.\nonumber
\end{align}

Note that the above conditions are not  linear constraints, which pose a great challenge to achieve $\mathscr{H}_2$ performance optimization model reduction and guarantee physical realizability for linear quantum systems. Next, we consider how to convert the nonlinear constraints into LMI constraints.

First, we present  necessary conditions for the solution of the optimization problem \eqref{opt}.

 \textbf{Theorem 1.} For the original model \eqref{dx2} with physical realizability conditions \eqref{real}, \eqref{real2} and \eqref{real3}, if there exist definite matrices $P$ and $Q$ that satisfy the Lyapunov equations \eqref{P} and \eqref{Q} and the matrices \( A_r \), \( B_r \), and \( C_r \) in the reduced model \eqref{dxr} are of the form
  \begin{align}
  A_r&= TAV, \label{Ar}\\
  B_r&=TB,\label{Br}\\
   C_r&=CV,\label{Cr}
\end{align}
 where $T=-Q_3^{-1}Q_2^T$ and $V=P_2P_3^{-1}$ satisfy
  \begin{align}
& TV=I,\label{TVI}\\
 &\mathbb J_n T^T=V \mathbb J_r,\label{TVJ}
\end{align}
  then \( A_r \), \( B_r \), and \( C_r \) are a solution to the optimization problem \eqref{opt} and the reduced model \eqref{dxr} is  physically realizable.

  \textbf{Proof}
  This proof can be divided into two parts. In the first part, we prove that \eqref{Ar}, \eqref{TVI} and \eqref{TVJ} are necessary conditions for the $\mathscr{H}_2$ optimization problem \eqref{opt}. In the second part, we establish that under the condition \eqref{TVJ}, the reduced model is physically realizable.

Part 1:
Supposing that $\nu$ is an arbitrary element in $A_r$, $B_r$ or $C_r$, the derivative of the cost function $J= \text{tr}(\hat CP\hat C^T)$ with respect to $\nu$ is calculated as
   \begin{align}
\frac{\partial J}{\partial \nu}=\mathrm{tr}\left(\frac{\partial P}{\partial \nu}\hat C^T\hat C\right)+\mathrm{tr}\left(\frac{\partial \hat C^T\hat C}{\partial \nu}P\right).\label{25-0}
\end{align}
Due to the difficulty in the calculation of $\frac{\partial P}{\partial \nu}$, we first utilize Lyapunov equation \eqref{Q} to rewrite Eq. \eqref{25-0}; i,e.,
   \begin{align}\label{21J}
\frac{\partial J}{\partial \nu}=-2\mathrm{tr}\left(\frac{\partial P}{\partial \nu}Q\hat A\right)+\mathrm{tr}\left(P\frac{\partial\hat  C^T\hat C}{\partial\nu}\right).
\end{align}
We then take the derivative of \eqref{P} with respect to $\nu$ to obtain
   \begin{align}\label{25b}
\frac{\partial \hat A}{\partial \nu}P+\hat A\frac{\partial P}{\partial \nu}+\frac{\partial P}{\partial \nu}\hat A^T+P\frac{\partial \hat A^T}{\partial \nu}+\frac{\partial \hat B\hat B^T}{\partial \nu}=0.
\end{align}
Post-multiplying  \eqref{25b} by $Q$ and taking the trace we obtain
   \begin{align} \label{23C}
2\mathrm{tr}\left(\frac{\partial \hat A}{\partial \nu}PQ\right)+2\mathrm{tr}\left(\frac{\partial P}{\partial \nu}Q\hat A\right)+\mathrm{tr}\left(\frac{\partial \hat B\hat B^T}{\partial \nu}Q\right)=0.
\end{align}
Hence, with \eqref{21J} and \eqref{23C}, Eq. \eqref{25-0} can be rewritten as
   \begin{align}
\frac{\partial J}{\partial \nu}=
2\mathrm{tr}\left(\frac{\partial \hat A}{\partial \nu}PQ\right)+\mathrm{tr}\left(\frac{\partial \hat B\hat B^T}{\partial \nu}Q\right)+\mathrm{tr}\left(\frac{\partial \hat C^T\hat C}{\partial \nu}P\right).
\end{align}
When $\nu$ is an arbitrary element of the matrix $C_r$, we have
   \begin{align}
0&=\frac{\partial J}{\partial \nu}=
\mathrm{tr}\left(\frac{\partial \hat C^T\hat C}{\partial \nu}P\right)\nonumber\\
&=\mathrm{tr}\left( \left[\begin{array}{cc}0&-C^T \frac{\partial C_r}{\partial \nu}\\-\frac{\partial C^T_r}{\partial \nu}C &-C^T \frac{\partial C_r^T}{\partial \nu}C_r+C^T_r\frac{\partial C_r}{\partial \nu}\\\end{array}\right]\left[\begin{array}{cc}P_1&P_2  \\P_2^T &P_3\\\end{array}\right]\right)\nonumber\\
&=\mathrm{tr}\left(-C^T\frac{\partial C_r}{\partial \nu}P^T_2\right)\nonumber\\
&+\mathrm{tr}\left(-\frac{\partial C^T_r}{\partial \nu}CP_2+\frac{\partial C^T_r}{\partial \nu}C_rP_3+C^T_r\frac{\partial C_r}{\partial \nu}P_3\right)\nonumber\\
&=2\mathrm{tr}\left(\frac{\partial C^T_r}{\partial \nu}(C_rP_3-CP_2)\right).
\end{align}
Since $\nu$ is an arbitrary element, we have $C_rP_3-CP_2=0$.
We arrive at
   \begin{align}
   C_r=CP_2P_3^{-1},
\end{align}
and the condition \eqref{Cr} has been established.

Similarly, when $\nu$ represents an arbitrary element of the matrix $B_r$, we  obtain
   \begin{align}
0&=\frac{\partial J}{\partial \nu}=
\mathrm{tr}\left(\frac{\partial \hat B\hat B^T}{\partial \nu}Q\right)\nonumber\\
&=\mathrm{tr}\left(\left[\begin{array}{cc}0&B \frac{\partial B_r^T}{\partial \nu}\\\frac{\partial B_r}{\partial \nu}B^T &\frac{\partial B_r}{\partial \nu}B_r^T+B_r\frac{\partial B_r^T}{\partial \nu}
\\\end{array}\right]\left[\begin{array}{cc}Q_1&Q_2  \\Q_2^T &Q_3\\\end{array}\right]\right)\nonumber\\
&=\mathrm{tr}\left(B\frac{\partial B^T_r}{\partial \nu}Q_2^T+\frac{\partial B_r}{\partial \nu}B^TQ_2+\frac{\partial B_r}{\partial \nu}B^T_rQ_3+B_r\frac{\partial B_r^T}{\partial \nu}Q_3\right)\nonumber\\
&=2\mathrm{tr}\left(\frac{\partial C^T_r}{\partial \nu}(Q_2^TB+Q_3B_r)\right).
\end{align}
Since $\nu$ is  arbitrary, we have $Q_2^TB+Q_3B_r=0$.
This yields the result
   \begin{align}
  B_r=-Q_3^{-1}Q_2^TB,\label{BrQ3}
\end{align}
and the condition \eqref{Br} has been obtained.

For any element $\nu$ in the matrix $A_r$, we have
  \begin{align}
0&=\frac{\partial J}{\partial \nu}=
2\mathrm{tr}\left(\frac{\partial \hat A}{\partial \nu}PQ\right)\nonumber\\
&=2\mathrm{tr}\left(\left[\begin{array}{cc}0&0  \\0 &\frac{\partial \hat A}{\partial \nu}\\\end{array}\right]\left[\begin{array}{cc}P_1&P_2  \\P_2^T &P_3\\\end{array}\right]\left[\begin{array}{cc}Q_1&Q_2  \\Q_2^T &Q_3\\\end{array}\right]\right)\nonumber\\
&=2\mathrm{tr}\left(\frac{\partial \hat A}{\partial \nu}(P_2^TQ_2+P_3Q_3)\right).
\end{align}

Since $\nu$ is an arbitrary entry, we have
  \begin{align}
P_2^TQ_2+P_3Q_3=0.\label{P2Q2}
\end{align}

The next step involves the calculation of $A_r$. To obtain $A_r$, we expand Eq. \eqref{P} and obtain
   \begin{align}
AP_1+P_1A^T+BB^T&=0, \\
AP_2+P_2A_r^T+BB^T_r&=0,\label{AP2}\\
A_rP_3+P_3A^T_r+B_rB^T_r&=0.\label{ArP3}
\end{align}

It can be observed that the equations \eqref{AP2} and \eqref{ArP3} are related to $A_r$.
Next, substituting \eqref{BrQ3} into \eqref{AP2} and \eqref{ArP3}, we have
   \begin{align}
AP_2+P_2A^T_r-BB^TQ_2Q^{-T}_3&=0\label{jk1},\\
A_rP_3+P_3A^T_r+Q^{-1}_3Q^T_2BB^TQ_2Q^{-T}_3&=0.\label{ArP33}
\end{align}

From Eq.  \eqref{jk1}, one has $Q^{-1}_3Q^T_2BB^TQ_2Q^{-T}_3=Q^{-1}_3Q^T_2AP_2+Q^{-1}_3Q^T_2P_2A^T_r$,
combined with \eqref{ArP33} we have
   \begin{align}
Q^{-1}_3Q^T_2AP_2+P_3A^T_r+(Q^{-1}_3Q^T_2P_2+P_3)A^T_r=0.\label{Q3Q236}
\end{align}

Due to Eq. \eqref{P2Q2}, we have
    \begin{align}
Q^{-1}_3Q^T_2P_2+P_3=0.
\end{align}

Thus, it can be deduced from Eq. \eqref{Q3Q236} that
 $A_r=-Q^{-1}_3Q^T_2AP_2P^{-1}_3$ and the condition \eqref{Ar} has been obtained. Pre- and post-multiplying \eqref{P2Q2} by $P_3^{-1}$ and $Q_3^{-1}$, respectively, we have
   \begin{align}
TV=-Q^{-1}_3Q^T_2P_2P^{-3}_3=I.
\end{align}

Therefore, we have proven the necessary conditions \eqref{Ar} and \eqref{TVI} for the reduced model to achieve the optimal $\mathscr{H}_2$ norm performance. Next, we will prove that the reduced model remains physically realizable.

Part 2:
To prove the physical realizability of the reduced system, we pre- and post-multiply by $T$ and $T^T$ on the physical realizability conditions for the original system \eqref{dx2} and also utilize \eqref{TVJ}, we get
\begin{align}
&TA\mathbb J_nT^T+T\mathbb J_nA^TT^T+TB\mathbb J_mB^TT^T\nonumber\\
&=TAV\mathbb J_r+\mathbb J_rV^TA^TT^T+B_r\mathbb J_mB_r^T\nonumber\\
&=A_r\mathbb J_r+\mathbb J_rA_r^T+B_r\mathbb J_mB_r^T=0.
\end{align}
 Next, by pre-multiplying the physical realizability condition of the original system \eqref{real2} by $T$, we arrive at the second condition for physical realizability for the reduced system \eqref{realr2}
\begin{align}
T\mathbb J_n C^T+TB\mathbb J_mD^T&=\mathbb J_rC^T_r+B_r\mathbb J_mD^T\nonumber\\
&=0.
\end{align}
Hence   $A_r= TAV$, $B_r=TB$, $C_r=CV$ and \eqref{DrD} can minimize the $\mathscr{H}_2$ norm and satisfy the  physical realizability conditions \eqref{realr} and \eqref{realr2}.
$\blacksquare$

  \textbf{Remark 1.} Theorem 1 presents necessary conditions for the optimal $\mathscr{H}_2$ reduction of linear quantum   systems, ensuring that the reduced model is physically realizable.
  Unfortunately,  the conditions in Theorem 1 are not easily solvable.
   This is because \(P\), \(Q\), \(\hat A\), \(\hat B\), and \(\hat C\) are matrices that need to be determined, resulting in the formation of nonlinear matrix equations \eqref{P}, \eqref{Q}, \eqref{TVI} and \eqref{TVJ} that are challenging to solve. In order to facilitate the solution process, it is necessary to transform these nonlinear matrix equations into linear matrix equations.
  We will  derive an equivalent form of Theorem 1, for which the constraints are  readily solvable.

\subsection{LMI based $\mathscr{H}_2$ Optimal Model Reduction Method}

To transform the conditions in Theorem 1 into a  solvable form,
we will transform the optimization problem \eqref{opt} with equality constraints to one with inequality constraints.

Firstly, for the Lyapunov equation \eqref{Q}, if we transform the solution \( Q \) to \( \hat{Q} = Q + \epsilon_Q \), where \( \epsilon_Q \) is a small positive definite matrix with appropriate dimensions, the above equation becomes \( \hat A^T\hat{Q} + \hat{Q}\hat A + \hat C^T\hat C < 0 \). Consequently, we have \( \text{tr}(\hat B^T Q \hat B) < \text{tr}(\hat B^T \hat{Q} \hat B) \).
So we consider a positive real constant $\gamma$ as an upper bound of $||\Xi_G-\Xi_{G_r}||_{2}$.
 This leads us to formulate the following optimization problem.

For the stable linear quantum  system  \eqref{dx2} with the physical realizability conditions \eqref{real}, \eqref{real2} and \eqref{real3},  find a Hurwitz matrix $A_r$, real-valued matrices $B_r$, $C_r$ and positive definite matrices $\hat Q$ that minimize the square of the $\mathscr{H}_2$ norm \eqref{XiG} with  physical realizability conditions \eqref{realr} and \eqref{realr2}; i.e.,

\begin{align}
    & \min_{\hat{Q}, A_r, B_r, C_r} \gamma^2 \label{opt2} \\
    \text{s.t.} \quad
    & \hat{Q} = \begin{bmatrix}
        \hat{Q}_1 & \hat{Q}_2 \\
        \hat{Q}_2^T & \hat{Q}_3
    \end{bmatrix} > 0, \label{hatQ} \\
    & \hat A^T \hat{Q} + \hat{Q} \hat A + \hat C^T  \hat C < 0, \label{hatQAe} \\
    & \operatorname{tr}( \hat B^T \hat{Q}  \hat B)-\gamma^2 < 0, \label{hatQBe}\\
    &\text{Eqs.}\  \eqref{realr},\ \eqref{realr2}.\nonumber
\end{align}

It can then be observed that the upper bound on the $\mathscr{H}_2$ norm on the difference between the transfer functions of the original and the reduced models can be expressed as $||\Xi_G-\Xi_{G_r}||_{2}<\gamma$.
 Once $\gamma^2$ is minimized, then the $\mathscr{H}_2$ norm of the difference between the transfer functions of the original and reduced models is also minimized.

 Note that at this point, the Lyapunov equation \eqref{Q} in the optimization problem \eqref{opt} has been transformed into the inequality condition \eqref{hatQAe}.
  If we consider an equivalent transformation of the condition \eqref{hatQAe}  to obtain LMI conditions, we can derive equivalent conditions for \eqref{Q}.
 The main findings are as follows.

%
%
%

\textbf{Theorem 2.} For model \eqref{dx2} with physical realizability conditions \eqref{real}, \eqref{real2} and \eqref{real3}, if there exist a scalar $\gamma$, positive  symmetric  matrices $\hat Q_{11}\in\mathbb{R}^{2r\times2r}$, $\hat Q_{12}\in\mathbb{R}^{2r\times2(n-r)}$, $\hat Q_{22}\in\mathbb{R}^{2(n-r)\times2(n-r)}$, $\hat Q_{2}\in\mathbb{R}^{2n\times2r}$,
    $\hat Q_{3}\in\mathbb{R}^{2r\times2r}$ and
   $M_r\in\mathbb{R}^{2r\times2r}$  such that the following optimization problem is solvable
    \begin{align}
& \min_{\hat{Q}_1,\hat{Q}_2, \hat{Q}_3, M_r} \gamma^2 \label{opt3} \\
    \text{s.t.} \quad
    &\hat Q_1=\left[\begin{array}{cc}\hat Q_{11}&\hat Q_{12}  \\\hat Q_{12}^T &\hat Q_{22}\\\end{array}\right]>0,\label{Q1Q3Q}\\
     &\left[\begin{array}{cccc}\Gamma_{11} & \Gamma_{12}& \Gamma_{13}& \Gamma_{14}\\\#&\Gamma_{22} &\Gamma_{23} & \Gamma_{24}\\\#&\#&\Gamma_{33}& \Gamma_{34}\\\#&\#&\#& \Gamma_{44}\\\end{array}\right] <0,\label{36}
\\
&\mathrm{tr}\left(B^T\left[\begin{array}{cc}\hat Q_{11}+3M_r&\hat Q_{12}  \\\hat Q_{12}^T &\hat Q_{22}\\\end{array}\right]B\right)<\gamma^2,\label{38}
\\
&\hat Q_1\mathbb J_n \hat Q_2-\hat Q_2\mathbb J_r\hat Q_3=0, \label{39}\\
&\left[\begin{array}{cc}M_r&0 \\ 0&0\\\end{array}\right]-\hat Q_2 \hat Q_3 ^{-1}\hat Q_2^T=0, \label{40}
\end{align}
        with
        \begin{align}
  &\Gamma_{11}=\mathrm{He}(A^T_{11}\hat Q_{11}+A_{21}^T\hat Q^T_{12})+C_1^TC_1,\nonumber\\
  &\Gamma_{12}=A^T_{11}\hat Q_{12}+A_{21}^TQ_{12}^T+\hat Q_{11}A_{12}+\hat Q_{12}A_{22}+C_1^TC_2,\nonumber\\
  &\Gamma_{13}=A^T_{11}\hat Q_{11}+A_{21}\hat Q_{12}^T+M_rA_{11}-C^T_1C_1,\nonumber\\
  &\Gamma_{14}=A^T_{11}\hat Q_{12}+A^T_{21}\hat Q_{22}+M_rA_{12}-C^T_1C_2,\nonumber\\
  &\Gamma_{22}=\mathrm{He}(A_{12}^T\hat Q_{12}+A^T_{22}\hat Q_{22})+C_2^TC_2,\nonumber\\
  &\Gamma_{23}=A_{12}^T\hat Q_{11}+A^T_{22}\hat Q_{12}^T-C_2^TC_1,\nonumber\\
  &\Gamma_{24}=A_{12}^T\hat Q_{12}+A^T_{22}\hat Q_{22}-C_2^TC_2,\nonumber\\
  &\Gamma_{33}=\mathrm{He}(A^T_{11}\hat Q_{11}+A_{21}^T\hat Q^T_{12})+C_1^TC_1,\nonumber\\
  &\Gamma_{34}=A^T_{11}\hat Q_{12}+A_{21}^TQ_{12}^T+\hat Q_{11}A_{12}+\hat Q_{12}A_{22}+C_1^TC_2,\nonumber\\
  &\Gamma_{44}=\mathrm{He}(A_{12}^T\hat Q_{12}+A^T_{22}\hat Q_{22})+C_2^TC_2\nonumber,
  \end{align}
   then, the reduced model \eqref{dxr} minimizes the $\mathscr{H}_2$ norm of $\Xi_G-\Xi_{G_r}$
   with \begin{align}
  A_r= \hat TA\hat V, B_r=\hat TB, C_r=C\hat V,
\end{align}
where $\hat T=\hat Q_3^{-1}\hat Q_2^{T} $, $\hat V=\hat Q_1^{-1}\hat Q_2$. Also, the reduced model \eqref{dxr} is physically realizable.

\textbf{Proof.}
In this proof, we show how to transform the nonlinear matrix inequality conditions \eqref{hatQAe} and \eqref{hatQBe}  into the linear matrix inequality conditions \eqref{36} and \eqref{38}.
Firstly, we define the $2n\times 2n$ matrix $M=\left[\begin{array}{cc}M_r&0 \\ 0&0\\\end{array}\right] $.
According to the matrix inequality constraint \eqref{36}, we get
 \begin{align}
&\left[\begin{array}{cc}\Gamma_{11}&\Gamma_{12} \\ \# &\Gamma_{22}\\\end{array}\right]=A^T\hat Q_1+\hat Q_1A+C^TC,\\
&\left[\begin{array}{cc}\Gamma_{13}&\Gamma_{14} \\ \Gamma_{23} &\Gamma_{24}\\\end{array}\right]=A^T\hat Q_1+MA-C^TC,\\
&\left[\begin{array}{cc}\Gamma_{33}&\Gamma_{34} \\ \# &\Gamma_{44}\\\end{array}\right]=A^T\hat Q_1+\hat Q_1A+C^TC.
\end{align}
 Thus, inequality \eqref{36} is equivalent to
 \begin{align}\label{ATQhat}
\left[\begin{array}{cc}A^T\hat Q_1+\hat Q_1A+C^TC&A^T\hat Q_1+MA-C^TC \\ \# &A^T\hat Q_1+\hat Q_1A+C^TC\\\end{array}\right]<0,
\end{align}
which is an LMI constraint.

Next we show that \eqref{ATQhat} and \eqref{hatQAe} are equivalent constraints.
By pre- and post-multiplying  both sides of Eq. \eqref{ATQhat} with the matrices $\left[\begin{array}{cc}I&0 \\ 0&\hat V^T\\\end{array}\right]$ and $\left[\begin{array}{cc}I&0 \\ 0&\hat V\\\end{array}\right]$ respectively, we obtain

\begin{align*}
&\left[\begin{array}{cc}I&0 \\ 0&\hat V^T\\\end{array}\right] \left[\begin{array}{c}
A^T\hat Q_1+\hat Q_1A+C^TC \\
 \# \\
\end{array}\right.\\
&\quad\ \left.\begin{array}{c}
A^T\hat Q_1+MA-C^TC   \\
A^T\hat Q_1+\hat Q_1A+C^TC  \\
\end{array}\right]\left[\begin{array}{cc}I&0 \\ 0&\hat V\\\end{array}\right]\nonumber\\
=&\left[\begin{array}{cc}I&0 \\ 0&\hat Q_2^T\hat Q_1^{-1}\\\end{array}\right] \left[\begin{array}{c}
A^T\hat Q_1+\hat Q_1A+C^TC \\
 \# \\
\end{array}\right.\\
&\quad\ \left.\begin{array}{c}
A^T\hat Q_1+MA-C^TC   \\
A^T\hat Q_1+\hat Q_1A+C^TC  \\
\end{array}\right]\left[\begin{array}{cc}I&0 \\ 0&\hat Q_1^{-1}\hat Q_2\\\end{array}\right]\nonumber\\
%
%
&= \left[\begin{array}{c}
A^T\hat Q_1+\hat Q_1A+C^TC \\
 \# \\
\end{array}\right.\\
&\quad\ \left.\begin{array}{c}
A^T\hat Q_2+MA\hat Q_1^{-1}\hat Q_2-C^TC\hat Q_1^{-1}\hat Q_2    \\
\hat Q^T_2\hat Q_1^{-1}A^T\hat Q_2+\hat Q_2^TA\hat Q_1^{-1}\hat Q_2+\hat Q^T_2\hat Q_1^{-1}C^TC\hat Q_1^{-1}\hat Q_2  \\
\end{array}\right]\nonumber\\
&= \left[\begin{array}{c}
A^T\hat Q_1+\hat Q_1A+C^TC \\
 \# \\
\end{array}\right.\\
&\quad\ \left.\begin{array}{c}
A^T\hat Q_2+\hat Q_2\hat Q_3^{-1}\hat Q_2^TA\hat Q_1^{-1}\hat Q_2-C^TC\hat Q_1^{-1}\hat Q_2    \\
\hat Q^T_2\hat Q_1^{-1}A^T\hat Q_2+\hat Q_2^TA\hat Q_1^{-1}\hat Q_2+\hat Q^T_2\hat Q_1^{-1}C^TC\hat Q_1^{-1}\hat Q_2  \\
\end{array}\right]\nonumber\\
&=\left[\begin{array}{cc}A^T\hat Q_1+\hat Q_1A+C^TC&A^T\hat Q_2+\hat Q_2A_r-C^TC_r \\ \# &A_r^T\hat Q_3+\hat Q_3A_r+C_r^TC_r\\\end{array}\right]\nonumber\\
&=\hat A^T \hat{Q} + \hat{Q} \hat A + \hat C^T  \hat C<0.
\end{align*}

 Thus, we obtain the matrix inequality constraint \eqref{hatQAe}.
Up to this point, we have proved that the condition \eqref{hatQAe} is equivalent to the condition \eqref{36}.

Further, we show that \eqref{hatQBe} and \eqref{38} are equivalent.
To eliminate the nonlinear term $\hat B^T \hat{Q}  \hat B$ in \eqref{hatQBe}, we decompose the matrix \(\hat Q=\left[\begin{array}{cc}\hat Q_1&\hat Q_2  \\ \hat Q_2^T &\hat Q_3\\\end{array}\right]\) and obtain
\begin{align}\label{BeQ}
&\mathrm{tr}(\hat B^T\hat Q\hat B)
=\mathrm{tr}\left(\left[\begin{array}{cc}B^T&B^T_r \\\end{array}\right]\left[\begin{array}{cc}\hat Q_1&\hat Q_2  \\\hat Q_2^T &\hat Q_3\\\end{array}\right]\left[\begin{array}{c}B\\B_r \\\end{array}\right]\right)\nonumber\\
&=\mathrm{tr}\left(\left[\begin{array}{cc}B^T&B^T\hat Q_2\hat Q_3^{-1} \\\end{array}\right]\left[\begin{array}{cc}\hat Q_1&\hat Q_2  \\ \hat Q_2^T &\hat Q_3\\\end{array}\right]\left[\begin{array}{c}B\\\hat Q_3^{-1}\hat Q_2^TB \\\end{array}\right]\right)\nonumber\\
&=\mathrm{tr}\left(B^T\left[\begin{array}{cc}\hat Q_{11}+3M_r&\hat Q_{12}  \\\hat Q_{12}^T &\hat Q_{22}\\\end{array}\right]B\right).\nonumber
\end{align}

At this point, the matrix inequality condition \eqref{hatQBe} has been transformed into the linear matrix inequality condition \eqref{38}.
From the physical realizability
condition \eqref{TVJ} in Theorem 1,  we have
\begin{equation}\label{566}
\mathbb J_n \hat T^T-\hat V \mathbb J_r=\mathbb J_n\hat Q_2\hat Q_3^{-1}-\hat Q_1^{-1} \hat Q_2 \mathbb J_r=0.
\end{equation}
By pre- and post-multiplying $\hat Q_1$ and $\hat Q_3$ on \eqref{566},
\eqref{39} can be easily derived.
This ends the proof of the Theorem.
$\blacksquare$
%

As proved by Theorem 2, we have transformed the nonlinear matrix inequalities \eqref{hatQAe} and \eqref{hatQBe}  into LMIs and nonlinear matrix equations, such that the optimization problem can be solved by an LMI approach.
According to Theorem 2, there is still an error $||\Xi_G-\Xi_{G_r}||_{2}$ between the reduced model \eqref{dxr} obtained by applying Theorem 2 and the original model \eqref{dx2}.

Although Eqs. \eqref{Q1Q3Q}-\eqref{38} in Theorem 2 are LMIs, \eqref{39} and \eqref{40} are nonlinear matrix equations that are difficult to solve.
In order to facilitate the solution process, we will  employ a lifting variables approach to convert the nonlinear terms in Eqs. \eqref{39} and \eqref{40} to linear terms, where additional variables and auxiliary equation constraints should be added \cite{42}.

\subsection{Matrix Lifting Method for Model Reduction}


The matrix lifting method is a technique for addressing nonlinear constrained optimization problems, serving as an extension of variable lifting methods \cite{42}. By introducing auxiliary matrix variables, this approach transforms complex nonlinear constraints into linear matrix constraints, thereby enabling solutions within a convex optimization framework.
  In the following, we linearize the conditions \eqref{39} and \eqref{40} by introducing appropriate matrix lifting variables and related equation constraints.

Here, we aim to preserve the matrix structure of the problem and try to find suitable matrix lifting variants.
 Firstly, six matrix lifting variables $W_{1-6}$ are given as: $W_1=\hat Q_1 \mathbb J_n$, $W_2=\hat Q_2 \mathbb J_r$, $W_3=W_1\hat Q_2$, $W_4=W_2\hat Q_3$, $W_5=\hat Q_3^{-1}$ and $W_6=\hat Q_2^T$.
 Now, we define a matrix $\mathbf Z$ which is a $(9n+2r)\times(9n+2r)$ symmetric matrix.
  By replacing $\hat Q_1$, $\hat Q_2$, $\hat Q_3$, $M$, $\hat Q_3^{-1}$ and $\hat Q_2^T$ respectively with $Z_{x_1,1}$, $Z_{x_2,1}$, $Z_{x_3,1}$, $Z_{x_4,1}$, $Z_{x_5,1}$ and $Z_{x_6,1}$ where
  \begin{align}
Z_{i,j}=[\mathbf Z_{kl}]_{k=in+1:(i+1)n;l=jn+1:(j+1)n} ,
\end{align}
   the matrix $\mathbf Z$ satisfies the following constraints
\begin{align}
Z_{0,0}-I_n &=0,\label{64}\\
  Z_{x_1,1}-Z_{1,x_1}&=0 ,\\
  Z_{x_3,1}-Z_{1,x_3}&=0,\\
  Z_{x_4,1}-Z_{1,x_4}&=0 ,\\
  Z_{v_1,1}-Z_{x_1,1} \mathbb J_n& =0,\\
  Z_{v_2,1}-Z_{x_2,1} \mathbb J_r &=0,\\
  Z_{v_3,1}-Z_{v_1,x_2}&=0, \\
  Z_{v_4,1}-Z_{v_2,x_3}&=0,\\
  Z_{v_5,1}-Z_{x_2,x_5}&=0,\\
    Z_{v_6,1}-Z_{v_5,x_6}&=0,\\
    I_r-Z_{x_3,x_5}&=0,\label{74}
\end{align}
where
\begin{align}
Z_{a,b}=Z_{a,1}Z_{b,1},\label{zab}
\end{align}
 with $a,b\in\{v_1,...v_6\}\cup\{x_1,...,x_4\}$.
 The LMI constraints in Theorem 2 can be expressed in terms of $\mathbf Z$ by replacing \(\hat Q_1\), $\hat Q_2$, $\hat Q_3$ and \(M\) with $Z_{x_1,1},Z_{x_2,1},Z_{x_3,1},Z_{x_4,1}$ respectively, and Eqs. \eqref{Q1Q3Q}-\eqref{38} can be rewritten as
 \begin{align}
   & Z_{x_1,1},Z_{x_3,1}>0,  \label{135}\\
     &\left[\begin{array}{cc}A^TZ_{x_1,1}+Z_{x_1,1}A+C^TC&A^TZ_{x_1,1}+Z_{x_4,1}A-C^TC \\ \# &A^TZ_{x_1,1}+Z_{x_1,1}A+C^TC\\\end{array}\right]\nonumber\\&<0, \label{136}
\\
&\mathrm{tr}(B^T(Z_{x_1,1}+3Z_{x_4,1})B)<\gamma^2 ,\label{138}
\end{align}
respectively.
Next, we consider the transformation of the nonlinear equality matrix constraint \eqref{39} into linear matrix equality constraints.
From \eqref{zab}  one can easily obtain
  \begin{align}
Z_{v_3,1}=Z_{v_1,x_2}=Z_{v_1,1}Z_{x_2,1}=Z_{x_1,1}\mathbb J_nZ_{x_2,1}=\hat Q_1\mathbb J_n \hat Q_2,\\
Z_{v_4,1}=Z_{v_2,x_3}=Z_{v_2,1}Z_{x_3,1}=Z_{x_2,1}\mathbb J_rZ_{x_3,1}=\hat Q_2\mathbb J_r\hat Q_3,
\end{align}
 and the physical realizability constraint \eqref{39} can be replaced by
\begin{align}
  Z_{v_3,1}-Z_{v_4,1}&=0,\label{81}\\
   \textbf Z&\geq0.
\end{align}

Similarly, the nonlinear constraint \eqref{40} can be rewritten as
\begin{align}
  Z_{x_4,1}-Z_{v_5,v_6}=0.\label{83}
\end{align}

 This leads us to reformulate the optimization problem \eqref{opt3}.

For linear quantum  system  \eqref{dx2} with physical realizability conditions \eqref{real}, \eqref{real2} and \eqref{real3},  find a matrix $\mathbf Z$ that minimize the square of the $\mathscr{H}_2$ norm \eqref{XiG} with  physical realizability conditions \eqref{realr} and \eqref{realr2}; i.e.,
\begin{align}
    & \min_{\mathbf Z} \gamma^2\label{opt100} \\
    \text{s.t.} \quad &\text{Eqs.}\ \eqref{64}-\eqref{74}, \eqref{135}-\eqref{138}, \eqref{81}-\eqref{83}.\nonumber
\end{align}

As noted in Ref. \cite{18}, solvers for LMI problems with rank constraints cannot guarantee convergence from arbitrary starting points. Therefore, it is crucial to use a heuristic method to select the initial points for these algorithms. For a given \( \gamma  > 0 \), in order to solve the problem quickly, we suggest solving the LMIs \eqref{135}, \eqref{136} and \eqref{138} to obtain \(\hat Q_1\), $\hat Q_2$, $\hat Q_3$ and \(M\).
We set $Z_0=V_0V_0^T$ as a heuristic starting point with
\begin{align}
V_0=&[I_n^T,\hat Q_1^T,\hat Q_2^T,\begin{bmatrix}
        \hat{Q}_3  \\
        0
    \end{bmatrix}^T,M^T,\begin{bmatrix}
        \hat{Q}_3^{-1}  \\
        0
    \end{bmatrix}^T,\hat Q_2^T,(\hat Q_1\mathbb J_n)^T,\nonumber\\
  &(\hat Q_2\mathbb J_r)^T,(\hat Q_1\mathbb J_n\hat Q_2)^T,(\hat Q_2\mathbb J_r\hat Q_3)^T,\hat Q_2\hat Q_3^{-1},\hat Q_2\hat Q_3^{-1}\hat Q_2^T]^T . \label{164}
\end{align}

Therefore, the optimization problem \eqref{opt100} can be solved by an algorithm similar to that proposed in \cite{13} which is based on LMIRank \cite{50}, SeDuMi \cite{51} and Yalmip \cite{52}.

In general, reduced order models are not unique. Next, we  give an alternative model reduction method from another perspective.

\subsection{An Extension of Optimal Model Reduction Method}

It's worth noting that earlier, we have utilized $||\Xi_G-\Xi_{G_r}||^2_{2}=\operatorname{tr}(\hat B^TQ\hat B)$ to solve the optimization problem \eqref{opt2}.
According to \eqref{XiG}, by employing the equation $||\Xi_G-\Xi_{G_r}||^2_{2}=\operatorname{tr}(\hat CP\hat C^T)$, we can also derive a new optimization problem equivalent to \eqref{opt}, which can be stated as follows.

For the linear quantum  system  \eqref{dx2} satisfying physical realizability conditions \eqref{real}, \eqref{real2} and \eqref{real3},  find Hurwitz matrix $A_r$, real-valued matrices $B_r$, $C_r$ and positive definite matrices $\hat P$ that minimize the square of the $\mathscr{H}_2$ norm \eqref{XiG} satisfying the  physical realizability conditions \eqref{realr} and \eqref{realr2}; i.e.,
\begin{align}
    & \min_{\hat{P}, A_r, B_r, C_r} \gamma^2 \label{opt4} \\
    \text{s.t.} \quad
    & \hat{P} = \begin{bmatrix}
        \hat{P}_1 & \hat{P}_2 \\
        \hat{P}_2^T & \hat{P}_3
    \end{bmatrix} > 0, \label{hatP} \\
    & \hat A \hat{P} + \hat{P}  \hat A^T +  \hat B  \hat B^T < 0, \label{AehatP} \\
    & \operatorname{tr}( \hat C\hat P \hat C^T) -\gamma^2< 0, \label{CehatP}
    \\
    &\text{Eqs.}\  \eqref{realr},\ \eqref{realr2}.\nonumber
\end{align}

Similarly, the main findings are as follows.

 \textbf{Theorem 3.} For model \eqref{dx2}, if there exist a scalar $\gamma$, positive  symmetric  matrices $\hat P_{11}\in\mathbb{R}^{2r\times2r}$, $\hat P_{12}\in\mathbb{R}^{2r\times2(n-r)}$, $\hat P_{22}\in\mathbb{R}^{2(n-r)\times2(n-r)}$, $\hat P_{2}\in\mathbb{R}^{2n\times2r}$,
    $\hat P_{3}\in\mathbb{R}^{2r\times2r}$ and
   $N_r\in\mathbb{R}^{2r\times2r}$   such that the following optimization problem is solvable
    \begin{align}
& \min_{\hat{P_1},\hat{P_2}, \hat{P_3}, N_r} \gamma^2 \label{opt4} \\
    \text{s.t.} \quad
    &\hat P_1=\left[\begin{array}{cc}\hat P_{11}&\hat P_{12}  \\ \hat P_{12}^T &\hat P_{22}\\\end{array}\right], \hat P_3, \hat{P}, N_r>0,\label{th31}\\
     &\Psi=\left[\begin{array}{cccc}\Psi_{11} & \Psi_{12}& \Psi_{13}& \Psi_{14}\\\#&\Psi_{22} &\Psi_{23} & \Psi_{24}\\\#&\#&\Psi_{33}& \Psi_{34}\\\#&\#&\#& \Psi_{44}\\\end{array}\right] <0,\label{th32}
\\
&\mathrm{tr}\left(C \left[\begin{array}{cc}\hat P_{11} - N_r&\hat P_{12}  \\\hat P_{12}^T &\hat P_{22}\\\end{array}\right] C ^T\right)<\gamma^2,\label{th33}
\\
& \hat P_1 \mathbb J_n\hat P_2-\hat P_2\mathbb J_r\hat P_3=0 \label{PJ}\\
& \left[\begin{array}{cc}N_r&0 \\ 0&0\\\end{array}\right]-\hat P_2\hat P_3^{-1}\hat P_2^T=0 \label{Nr}
\end{align}
        with
        \begin{align}
  &\Psi_{11}=\mathrm{He}(A_{11}\hat P_{11}+A_{12}\hat P^T_{12})+B_1B_1^T,\nonumber\\
  &\Psi_{12}=A_{11}\hat P_{12}+A_{12}P_{22}+\hat P_{11}A_{21}^T+\hat P_{12}A_{22}^T+B_1B_2^T,\nonumber\\
  &\Psi_{13}=A_{11}\hat P_{11}+A_{12}\hat P_{12}+N_rA_{11}^T+B_1B_1^T,\nonumber\\
  &\Psi_{14}=A_{11}\hat P_{12}+A_{12}\hat P_{22}+N_rA_{21}^T+B_1B_2^T,\nonumber\\
  &\Psi_{22}=\mathrm{He}(A_{21}\hat P_{12}+A_{22}\hat P_{22})+B_2B_2^T,\nonumber\\
  &\Psi_{23}=A_{21}\hat P_{11}+A_{22}\hat P_{12}^T+B_2B_1^T,\nonumber\\
  &\Psi_{24}=A_{21}\hat P_{12}+A_{22}\hat P_{22}+B_2B_2^T,\nonumber\\
  &\Psi_{33}=\mathrm{He}(A_{11}\hat P_{11}+A_{12}\hat P^T_{12})+B_1B_1^T,\nonumber\\
  &\Psi_{34}=A_{11}\hat P_{12}+A_{12}P_{22}+\hat P_{11}A_{21}^T+\hat P_{12}A_{22}^T+B_1B_2^T,\nonumber\\
  &\Psi_{44}=\mathrm{He}(A_{21}\hat P_{12}+A_{22}\hat P_{22})+B_2B_2^T\nonumber.
  \end{align}
then, the reduced model \eqref{dxr} minimizes the $\mathscr{H}_2$ norm of $||\Xi_G-\Xi_{G_r}||_2$
   with \begin{align}
  A_r= \tilde TA\tilde V, B_r=\tilde TB, C_r=C\tilde V,
\end{align}
where $\tilde T=\hat P_2^T\hat P_1^{-1} $, $\tilde V=\hat P_2\hat P_3^{-1}$.

\textbf{Proof.}
The proof follows a similar approach to that of Theorem 2 and is thus omitted. $\blacksquare$

Next, we can linearize the nonlinear matrix condition \eqref{PJ} by matrix lifting variables.
The procedures here are similar to those in the previous subsection and thus are omitted here.

\section{Model Reduction for Completely Passive  Linear Quantum Systems}

A completely passive linear quantum system is a special type of linear quantum system characterized by the fact that all system elements are passive. This means the system does not gain energy  but can only dissipate or conserve energy~\cite{44}. Compared to general linear quantum systems, the physical realizability conditions of these systems  tend to be different~\cite{45}.
In this section, we present the model reduction method for passive linear quantum systems.

\subsection{Problem formulation for Passive Completely Linear Quantum Systems}

A class of passive linear quantum systems with annihilation operators can be described  by quantum stochastic differential equations (QSDEs)~\cite{45}
\begin{align}\label{dx}
&da(t)=Fa(t)dt+Gd\mathcal{A}(t) \nonumber\\
&d\mathcal Y(t)=Ha(t)dt+Kd\mathcal{A}(t) ,
\end{align}
where $F\in\mathbb{C}^{n\times n}$, $G\in\mathbb{C}^{n\times m}$, $H\in\mathbb{C}^{l\times n}$ and $K\in\mathbb{C}^{l\times m}$.
Here, $a(t)=[a_1(t),...,a_n(t)]^T$ is a vector of annihilation operators.
$\mathcal Y(t)$ and $\mathcal A(t)$ are the same as those in Section \ref{Problem}.

The system in the form of Eq. \eqref{dx} can also be written in a quadrature representation as follows
\begin{align}\label{dx22}
&dx(t)=Ax(t)dt+Bdw(t),\nonumber\\
&dy(t)=Cx(t)dt+Dd w(t),
\end{align}
where $A\in \mathbb R^{2n\times2n}$, $B\in \mathbb R^{2n\times2m}$, $C\in \mathbb R^{2m\times2n}$, $D\in \mathbb R^{2l\times2m}$. $x(t)$, $ w(t)$ and $y(t)$ are the same as those in Section \ref{Problem}.
It should be noted that matrices $A$, $B$, $C$ and $D$ can respectively be written as block matrices containing submatrices of $2\times2$ with dimensions $n\times n$, $n\times m$, $n\times l$ and $m\times l$ respectively. Thus, the block submatrix at the $s$-th row and $t$-th column can be written in the following form
\begin{align}
&A_{s,t}=\frac{1}{2}\left[\begin{array}{cc}F_{s,t}+F_{s,t}^*&i(F_{s,t}-F^*_{s,t})  \\-i(F_{s,t}-F_{s,t}^*) &F_{s,t}+F_{s,t}^*\\\end{array}\right],\nonumber\\
&B_{s,t}=\frac{1}{2}\left[\begin{array}{cc}G_{s,t}+G_{s,t}^*&i(G_{s,t}-G_{s,t}^*)  \\-i(G_{s,t}-G_{s,t}^*) &G_{s,t}+G_{s,t}^*\\\end{array}\right],\nonumber\\
&C_{s,t}=\frac{1}{2}\left[\begin{array}{cc}H_{s,t}+H_{s,t}^*&i(H_{s,t}-H_{s,t}^*)  \\-i(H_{s,t}-H_{s,t}^*) &H_{s,t}+H_{s,t}^*\\\end{array}\right],\nonumber\\
&D_{s,t}=\frac{1}{2}\left[\begin{array}{cc}K_{s,t}+K_{s,t}^*&i(K_{s,t}-K_{s,t}^*)  \\-i(K_{s,t}-K_{s,t}^*) &K_{s,t}+K_{s,t}^*\\\end{array}\right].\nonumber
\end{align}

 Different from the active linear quantum system \eqref{dx2}, the passive linear quantum system \eqref{dx22} satisfies the following  physically realizability conditions~\cite{4}
\begin{align}
A+A^T+BB^T&=0,\label{real21}\\
B&=-C^T,\label{real22}\\
D&=I.\label{real23}
\end{align}

It is notable that when the condition \eqref{real23} is satisfied, we have $l=m$.
Similar to Section \ref{Problem}, given the passive linear quantum system \eqref{dx22}, we will look for a reduced model in the form of Eq. \eqref{dxr}   satisfying the physically realizability conditions
 \begin{align}
A_r+A_r^T+B_rB_r^T&=0,\label{realr21}\\
B_r&=-C_r^T,\label{realr22}\\
D_r&=I.\label{realr23}
\end{align}
 such that $||\Xi_G(s)-\Xi_{G_r}(s)||_2$ is  minimized, where transfer functions $\Xi_G(s)$ and $\Xi_{G_r}(s)$ are in the same form of active linear quantum systems.
 It is notable that condition \eqref{realr23}  is naturally satisfied since $D_r=D$.

\subsection{Optimization Framework for Passive Model Reduction}

The challenge of model reduction for passive quantum systems lies in identifying a model \eqref{dxr} that minimizes the $\mathscr{H}_2$ norm difference between the transfer functions of the original and the reduced models while ensuring  physical realizability.
 The next step involves the derivation of the reduced model \eqref{dxr}.
In order to evaluate the difference between of the original model \eqref{dx2} and the reduced model \eqref{dxr}, we construct an augmented system, which can be obtained as \eqref{dx3}.

The stability of the augmented system \eqref{dx3} can be guaranteed due to the Hurwitz stability of both the system model \eqref{dx22} and the reduced model \eqref{dxr}. We can find positive definite matrices $\bar P$ and $\bar Q$ that satisfy Lyapunov equations
\begin{align}
    &\hat A \bar P + \bar P \hat A^T + \hat B \hat B^T   = 0, \label{barP} \\
    &\hat A^T \bar Q + \bar Q \hat A + \hat C^T \hat C  = 0. \label{barQ}
\end{align}


Since $\bar P$ and $\bar Q$ are symmetric $2(n+r)\times 2(n+r)$ dimensional matrices, they can be rewritten in a block matrix form as
\begin{align}
\bar P=\left[\begin{array}{cc}\bar P_1&\bar P_2  \\\bar P_2^T &\bar P_3\\\end{array}\right],
\bar Q=\left[\begin{array}{cc}\bar Q_1&\bar Q_2  \\\bar Q_2^T &\bar Q_3\\\end{array}\right],
\end{align}
 where $\bar P_1, \bar Q_1\in \mathbb R^{2n\times2n}$, $\bar P_2, \bar Q_2\in \mathbb R^{2r\times2n}$ and $\bar P_3, \bar Q_3\in \mathbb R^{2r\times2r}$.

In addition, the equivalent form for the square of the $\mathscr{H}_2$ norm  \eqref{H2} can be calculated as
   \begin{align}
||\Xi_G-\Xi_{G_s}||^2_2={\rm tr}(\bar B^T\bar Q\bar B)={\rm tr}(\bar C\bar P\bar C^T).\label{XiGs}
\end{align}

Therefore, we can transform the $\mathscr{H}_2$ model reduction problem into solving the following optimization problem.

For the passive linear quantum  system  \eqref{dx22} satisfying physical realizability conditions \eqref{real21} and \eqref{real22},  find a Hurwitz matrix $A_r$, real-valued matrices $B_r$, $C_r$ and positive definite matrices $\bar P$ and $\bar Q$ that minimize the square of the $\mathscr{H}_2$ norm \eqref{XiG}; i.e.,
\begin{align}
    & \min_{\bar P, \bar Q, A_r, B_r, C_r}{\rm tr}(\bar B^T\bar Q\bar B)={\rm tr}(\bar C\bar P\bar C^T) \label{optp} \\
     &~~~~~~~~\text{s.t.}\
     \text{Eqs.}\ \ \eqref{realr21},\ \eqref{realr22},\ \eqref{barP},\ \eqref{barQ}.\nonumber
\end{align}

Next, we present necessary conditions for the solution of this optimization problem \eqref{optp}.

\textbf{Theorem 4.} For the original model \eqref{dx22} satisfying the physical realizability conditions \eqref{real21} and \eqref{real22}, if there exist positive definite matrices  $\bar P$ and $\bar Q$ that satisfy Lyapunov equations \eqref{barP} and \eqref{barQ} and the matrices \( A_r \), \( B_r \), and \( C_r \) in the reduced model \eqref{dxr} are of the form
  \begin{align}\label{Ass}
  A_r= TAV, B_r=TB, C_r=CV,
\end{align}
 where  $T=-\bar Q_3^{-1}\bar Q_2^T$ and $V=\bar P_2\bar P_3^{-1}$ satisfy
  \begin{align}
 TV&=I,\label{TVI2}\\
 T&=V^T,\label{TVT}
\end{align}
  then \( A_r \), \( B_r \), and \( C_r \) are a solution to the optimization problem \eqref{optp} and the reduced model \eqref{dxr} is  physically realizable.

 \textbf{Proof.}
The proof of this theorem consists of two parts. The first part is similar to the first part of the proof of Theorem 1. By computing the derivative of the cost function, we can obtain that \eqref{Ass} are a solution to the optimization problem \eqref{optp} and Eq. \eqref{TVI2} has now been proved.

Additionally, based on \eqref{TVT}, we can deduce
\begin{align}
A_r+A_r^T+B_rB_r^T&=TAV+V^TA^TV+TBB^TV\nonumber\\
&=T(A+A^T+BB^T)V=0,\\
B_r&=TB=-V^TC^T=-C_r^T.
\end{align}
Up to this point, we have proved that the reduced model \eqref{dxr} is physically realizable.
$\blacksquare$

In contrast to the optimization problem for the active linear quantum system in Theorem 1,  the passive linear quantum system  has to satisfy another nonlinear equality constraint \eqref{TVT}.
The nonlinear inequality constraints \eqref{TVI2} and \eqref{TVT} in Theorem 4 are still difficult to solve.
We derive the following result to obtain more tractable constraint conditions.

\textbf{Theorem 5.} For the original model \eqref{dx22} satisfying the physical realizability conditions \eqref{real21} and \eqref{real22}, if there exist a scalar $\gamma$, positive  symmetric  matrices $\bar Q_{11}\in\mathbb{R}^{2r\times2r}$, $\bar Q_{12}\in\mathbb{R}^{2r\times2(n-r)}$, $\bar Q_{22}\in\mathbb{R}^{2(n-r)\times2(n-r)}$, $\bar Q_{2}\in\mathbb{R}^{2n\times2r}$,
    $\bar Q_{3}\in\mathbb{R}^{2r\times2r}$ and
   $\bar M_r\in\mathbb{R}^{2r\times2r}$  such that the following optimization problem is solvable
    \begin{align}
& \min_{\bar{Q}_1,\bar{Q}_2, \bar{Q}_3, \bar M_r} \gamma^2 \label{opt5} \\
    \text{s.t.} \quad
    &\bar Q_1=\left[\begin{array}{cc}\bar Q_{11}&\bar Q_{12}  \\\bar Q_{12}^T &\bar Q_{22}\\\end{array}\right]>0,\label{143}\\
     &\left[\begin{array}{cccc}\Gamma_{11} & \Gamma_{12}& \Gamma_{13}& \Gamma_{14}\\\#&\Gamma_{22} &\Gamma_{23} & \Gamma_{24}\\\#&\#&\Gamma_{33}& \Gamma_{34}\\\#&\#&\#& \Gamma_{44}\\\end{array}\right] <0,\label{144}
\\
&\mathrm{tr}\left(B^T\left[\begin{array}{cc}\bar Q_{11}+3\bar M_r&\bar Q_{12}  \\\bar Q_{12}^T &\bar Q_{22}\\\end{array}\right]B\right)<\gamma^2,\label{145}
\\
&\left[\begin{array}{cc}\bar M_r&0 \\ 0&0\\\end{array}\right]-\bar Q_2 \bar Q_3 ^{-1}\bar Q_2^T=0, \label{146}\\
& \bar Q_1\bar Q_2-\bar Q_2\bar Q_3=0,\label{Q1Q2}
\end{align}
        with
        \begin{align}
  &\Gamma_{11}=\mathrm{He}(A^T_{11}\bar Q_{11}+A_{21}^T\bar Q^T_{12})+C_1^TC_1,\nonumber\\
  &\Gamma_{12}=A^T_{11}\bar Q_{12}+A_{21}^TQ_{12}^T+\bar Q_{11}A_{12}+\bar Q_{12}A_{22}+C_1^TC_2,\nonumber\\
  &\Gamma_{13}=A^T_{11}\bar Q_{11}+A_{21}\bar Q_{12}^T+\bar M_rA_{11}-C^T_1C_1,\nonumber\\
  &\Gamma_{14}=A^T_{11}\bar Q_{12}+A^T_{21}\bar Q_{22}+\bar M_rA_{12}-C^T_1C_2,\nonumber\\
  &\Gamma_{22}=\mathrm{He}(A_{12}^T\bar Q_{12}+A^T_{22}\bar Q_{22})+C_2^TC_2,\nonumber\\
  &\Gamma_{23}=A_{12}^T\bar Q_{11}+A^T_{22}\bar Q_{12}^T-C_2^TC_1,\nonumber\\
  &\Gamma_{24}=A_{12}^T\bar Q_{12}+A^T_{22}\bar Q_{22}-C_2^TC_2,\nonumber\\
  &\Gamma_{33}=\mathrm{He}(A^T_{11}\bar Q_{11}+A_{21}^T\bar Q^T_{12})+C_1^TC_1,\nonumber\\
  &\Gamma_{34}=A^T_{11}\bar Q_{12}+A_{21}^TQ_{12}^T+\bar Q_{11}A_{12}+\bar Q_{12}A_{22}+C_1^TC_2,\nonumber\\
  &\Gamma_{44}=\mathrm{He}(A_{12}^T\bar Q_{12}+A^T_{22}\bar Q_{22})+C_2^TC_2\nonumber,
  \end{align}
   then, the  physically realizable reduced model \eqref{dxr} minimizes the $\mathscr{H}_2$ norm of $||\Xi_G-\Xi_{G_r}||$
   with \begin{align}
  A_r= \bar TA\bar V, B_r=\bar TB, C_r=C\bar V, \label{th5Ar}
\end{align}
 where  $\bar T=\bar Q_3^{-1}\bar Q_2^{T} $, $\bar V=\bar Q_1^{-1}\bar Q_2$.

 \textbf{Proof.}
The proof of this theorem consists of two parts. Inspired by the proof of Theorem 2, we  obtain that \eqref{143}-\eqref{146} are necessary conditions for \eqref{th5Ar} as a solution to the optimization problem \eqref{optp}.

 Next, by employing the condition \eqref{Q1Q2}, we have
 \begin{align}
A_r+A_r^T+B_rB_r^T&=\bar TA\bar V+\bar V^TA^T\bar V+\bar TBB^T\bar V\nonumber\\
&=\bar Q_3^{-1}\bar Q_2^{T}(A+A^T+BB^T)\bar Q_1^{-1}\bar Q_2\nonumber\\
&=0,
\end{align}
and
   \begin{align}
B_r=\bar TB=\bar Q_3^{-1}\bar Q_2^{T}B=-C_r^T.
\end{align}
 This ends the proof.
 $\blacksquare$

In order to solve the optimization problem in Theorem 5, we still adopt the method of matrix lifting.
The procedures outlined in this section closely resemble those presented in the preceding subsection and are therefore omitted here.

  \textbf{Theorem 6.} For the original model \eqref{dx22} satisfying the physical realizability conditions \eqref{real21} and \eqref{real22}, if there exist a scalar $\gamma$, positive  symmetric  matrices $\bar P_{11}\in\mathbb{R}^{2r\times2r}$, $\bar P_{12}\in\mathbb{R}^{2r\times2(n-r)}$, $\bar P_{22}\in\mathbb{R}^{2(n-r)\times2(n-r)}$, $\bar P_{2}\in\mathbb{R}^{2n\times2r}$,
    $\bar P_{3}\in\mathbb{R}^{2r\times2r}$ and
   $\bar N_r\in\mathbb{R}^{2r\times2r}$   such that the following optimization problem is solvable
    \begin{align}
& \min_{\bar{P_1},\bar{P_2}, \bar{P_3},\bar  N_r} \gamma^2 \label{opt6} \\
    \text{s.t.} \quad
    &\bar P_1=\left[\begin{array}{cc}\bar P_{11}&\bar P_{12}  \\ \bar P_{12}^T &\bar P_{22}\\\end{array}\right], \bar P_3, \bar{P}, \bar N_r>0,\label{th131}\\
     &\Psi=\left[\begin{array}{cccc}\Psi_{11} & \Psi_{12}& \Psi_{13}& \Psi_{14}\\\#&\Psi_{22} &\Psi_{23} & \Psi_{24}\\\#&\#&\Psi_{33}& \Psi_{34}\\\#&\#&\#& \Psi_{44}\\\end{array}\right] <0,\label{th132}
\\
&\mathrm{tr}\left(C \left[\begin{array}{cc}\bar P_{11} - \bar N_r&\bar P_{12}  \\\bar P_{12}^T &\bar P_{22}\\\end{array}\right] C ^T\right)<\gamma^2,\label{th133}
\\
& \left[\begin{array}{cc}\bar N_r&0 \\ 0&0\\\end{array}\right]-\bar P_2\bar P_3^{-1}\bar P_2^T=0 ,\label{Nr2}\\
& \bar P_1\bar P_2-\bar P_2\bar P_3=0,\label{P1P2}
\end{align}
        with
        \begin{align}
  &\Psi_{11}=\mathrm{He}(A_{11}\bar P_{11}+A_{12}\bar P^T_{12})+B_1B_1^T,\nonumber\\
  &\Psi_{12}=A_{11}\bar P_{12}+A_{12}P_{22}+\bar P_{11}A_{21}^T+\bar P_{12}A_{22}^T+B_1B_2^T,\nonumber\\
  &\Psi_{13}=A_{11}\bar P_{11}+A_{12}\bar P_{12}+\bar N_rA_{11}^T+B_1B_1^T,\nonumber\\
  &\Psi_{14}=A_{11}\bar P_{12}+A_{12}\bar P_{22}+\bar N_rA_{21}^T+B_1B_2^T,\nonumber\\
  &\Psi_{22}=\mathrm{He}(A_{21}\bar P_{12}+A_{22}\bar P_{22})+B_2B_2^T,\nonumber\\
  &\Psi_{23}=A_{21}\bar P_{11}+A_{22}\bar P_{12}^T+B_2B_1^T,\nonumber\\
  &\Psi_{24}=A_{21}\bar P_{12}+A_{22}\bar P_{22}+B_2B_2^T,\nonumber\\
  &\Psi_{33}=\mathrm{He}(A_{11}\bar P_{11}+A_{12}\bar P^T_{12})+B_1B_1^T,\nonumber\\
  &\Psi_{34}=A_{11}\bar P_{12}+A_{12}P_{22}+\bar P_{11}A_{21}^T+\bar P_{12}A_{22}^T+B_1B_2^T,\nonumber\\
  &\Psi_{44}=\mathrm{He}(A_{21}\bar P_{12}+A_{22}\bar P_{22})+B_2B_2^T\nonumber.
  \end{align}
then, the  physically realizable reduced model \eqref{dxr} minimizes the $\mathscr{H}_2$ norm of $||\Xi_G-\Xi_{G_r}||_2$
   with \begin{align}\label{th6Ar}
  A_r= \tilde TA\tilde V, B_r=\tilde TB, C_r=C\tilde V,
\end{align}
 where  $\tilde T=\bar P_2^T\bar P_1^{-1} $, $\tilde V=\bar P_2\bar P_3^{-1}$.

 \textbf{Proof.}
The proof of this theorem consists of two parts.
Firstly, the proof of Theorem 3 serves as inspiration, leading us to conclude that conditions \eqref{th131}-\eqref{Nr2} are necessary conditions for \eqref{th6Ar} to be a solution to the optimization problem \eqref{optp}.

Next, by employing the condition \eqref{P1P2},  the physical realizable conditions \eqref{realr21} and \eqref{realr22} are obtained. $\blacksquare$

We also use the matrix lifting method to solve Theorem 6, as the procedures are similar to those in the previous section and will not be repeated.

\section{Examples}

In this section, we give two examples of active and passive linear quantum systems to illustrate the effectiveness of the proposed methods.

\subsection{An Example of Active Systems}
Firstly, we consider an optomechanical system in Ref. \cite{39} for back-action evading position measurement. This system consists of an optical cavity with two movable mirrors, see \cite{40} and \cite{41}. For studies on control in such systems, refer to \cite{41,41.1,41.2}. The cavity is pumped by a strong coherent laser, and each mirror experiences radiation pressure and thermal noise.

This system includes three degrees of freedom: an oscillator inside the cavity, described by the quadratures \((q_1, p_1)\), and two mechanical oscillators from the motion of the mirrors, described by the quadratures \((q_2, p_2, q_3, p_3)\). The dynamics can be linearized around the steady-state mean amplitude of the cavity mode. The linear approximation is given in a quadrature form (1) with \(x = (q_1, p_1, q_2, p_2, q_3, p_3)^T\) and the following system matrices:

\begin{align}
A &= \begin{bmatrix}
-\frac{\kappa}{2} & 0 & 0 & 0 & 0 & 0 \\
0 & -\frac{\kappa}{2} & -\Gamma & 0 & 0 & 0 \\
0 & 0 & -\frac{\gamma}{2} & 0 & 0 & \Omega_b \\
-\Gamma & 0 & 0 & -\frac{\gamma}{2} & -\Omega_b & 0 \\
0 & 0 & 0 & \Omega_b & -\frac{\gamma}{2} & 0 \\
0 & 0 & -\Omega_b & 0 & 0 & -\frac{\gamma}{2}
\end{bmatrix},\\
B &= \text{diag}(\sqrt{\kappa} I_2, \sqrt{\gamma} I_4),\\
C& = [\sqrt{\kappa} I_2 \, 0_{2 \times 4}], \\
 D &= \begin{bmatrix}
-I_2 & 0_{2 \times 4}
\end{bmatrix}.
\end{align}

Here, \(\kappa > 0\) represents the cavity decay rate, \(\gamma > 0\) is the damping rate of the mechanical oscillators, \(\Gamma > 0\) denotes the optomechanical coupling rate (due to the interaction between the mirror degrees of freedom and the cavity mode via radiation pressure), and \(\Omega_b\) is the system half-bandwidth.
 Inputs 1-2 correspond to the laser field quadratures, while inputs 3-6 account for thermal fluctuations affecting the mirrors. This system is active because the coupling \(\Gamma\) induces a squeezing Hamiltonian.

 Consistent with Ref. \cite{14}, let \(\kappa = 2 \times 10^5\mathrm{Hz}\), \(\gamma = 100\mathrm{Hz}\), \(\Gamma = 7.0711 \times 10^4\mathrm{Hz}\), and \(\Omega = 10^4\mathrm{Hz}\). After calculating the real parts of the system's poles, it was verified that they are all with negative real parts, which means that the system is stable.
It is easily checked that this original system  fulfills the criteria for physical realizability as outlined in Eq. \eqref{real}. Our aim is to simplify this original system into a fourth-order system according to the $\mathscr{H}_2$ norm, which also satisfies the physical realizable conditions.

Based on Theorem 2, we have derived a reduced model whose system matrices are
\begin{align}
A_r& =
\begin{bmatrix}
-41449 & 7913 & -45389 & 5330 \\
-58081 & -35985 & -59739 & -43039 \\
-40045 & 11234 & -44146 & 9000 \\
-66094 & -38193 & -68208 & -46019
\end{bmatrix},\\
B_r& =
\begin{bmatrix}
287.18 & 16.83 & 6.43 & 0.45 & 5.95 & 0.22 \\
-27.62 & 267.76 & -0.73 & 5.99 & -1.13 & 5.97 \\
310.24 & -2.36 & 6.96 & 0.02 & 6.47 & -0.22 \\
-6.24 & 290.40 & -0.26 & 6.51 & -0.73 & 6.46
\end{bmatrix},\\
C_r &=
\begin{bmatrix}
267.76 & -16.83 & 290.40 & 2.36 \\
27.62 & 287.18 & 6.24 & 310.24
\end{bmatrix}.
\end{align}

From the system matrices of the reduced model, we can see that the reduced model is also stable.
 By calculating the $\mathscr{H}_2$ norm, we obtain $||\Xi_G-\Xi_{G_r}||_{2}=528.36$.

\begin{figure}[htbp]
\centering
\includegraphics[width=9cm]{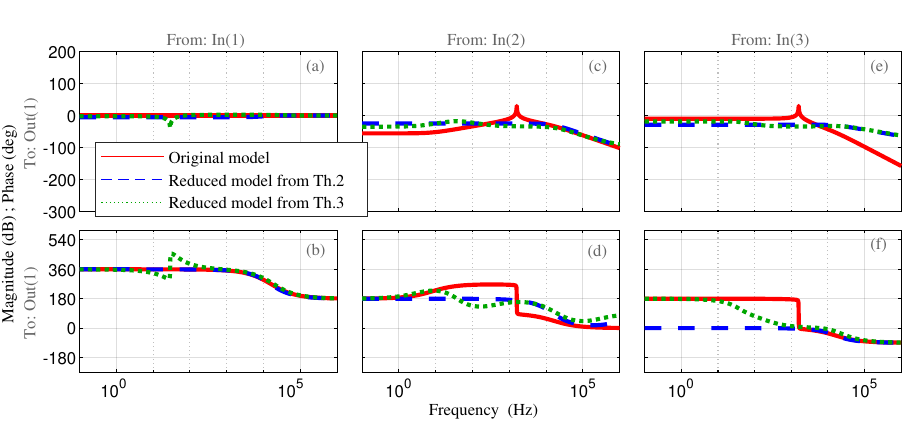}
\caption{Comparison of frequency responses between the original and reduced models for an active system}
\label{fig1.1}
\end{figure}

In Fig. \ref{fig1.1}, we compare the Bode diagrams of the model after reduction with those of the original model.
The red solid line and the blue dashed line show the frequency response at output 2 for the original system and the reduced system, respectively. Similar to Ref. \cite{14}, the responses from inputs 1, 4, and 5 are omitted due to dispersive coupling, as their transfer functions are zero.
 It can be seen from Fig. \ref{fig1.1} (a) and (b) that both systems have the same amplitude response at input 2, and the phase response of the reduced system is consistent with that of the original system at all frequencies.
In Fig. \ref{fig1.1} (c) and (d), the phase response of the reduced system matches that of the original system in the low frequency band, while in Fig. \ref{fig1.1} (e) and (f), the amplitude response of the reduced system is close to that of the original system in the low frequency band.
Compared with the interpolation projection reduction method in Ref. \cite{14}, the method in this paper not only has excellent fitting performance at the interpolation points in terms of frequency, but also achieves good results across the entire frequency range.
The advantage of the observed model reduction is that it significantly simplifies the original system while preserving its frequency response as much as possible.
 By maintaining a close match in both magnitude and phase responses at critical frequency ranges, the reduced model offers computational efficiency and ease of analysis. This is particularly beneficial in control system design and analysis, where simpler models can lead to more straightforward controller designs and faster simulations.


\subsection{An Example of Passive Systems}
We now explore a passive linear quantum system composed of three oscillators connected in series, designated as the model to be reduced.  The dynamics of this system follow the quantum Langevin equation

\begin{align}
\dot a_1&=(-i\omega_1-\frac{\kappa_1}{2})a_1-\sqrt{\kappa_1}b_{1,\mathrm{in}},\\
\dot a_2&=(-i\omega_2-\frac{\kappa_2}{2})a_2-\sqrt{\kappa_2}b_{2,\mathrm{in}},\\
\dot a_3&=(-i\omega_3-\frac{\kappa_3}{2})a_3-\sqrt{\kappa_3}b_{3,\mathrm{in}},\\
b_{1,\mathrm{out}}&=b_{1,\mathrm{in}}+\sqrt \kappa_1a_1,\\
b_{2,\mathrm{out}}&=b_{2,\mathrm{in}}+\sqrt \kappa_2a_2,\\
b_{3,\mathrm{out}}&=b_{3,\mathrm{in}}+\sqrt \kappa_3a_3,
\end{align}
where \(a_j\), \(b_{\text{j,in}}\) and $b_\text{j,out}$  denote the annihilation operator of the \(j_\text{th}\) harmonic oscillator, the input noise operator and the output operator, respectively. The frequencies of the three harmonic oscillators are set as \(\omega_1 = \omega_2 = 10\mathrm{Hz}\) and \(\omega_3 = 0.01\mathrm{Hz}\). The system's dissipation rates are specified as \(\kappa_1 = \kappa_2 = \kappa_3= 1\mathrm{Hz}\). Noting that the harmonic oscillators are connected in series, one can derive the input-output relationship accordingly
\begin{align}
&b_{2,\mathrm{in}}=b_{1,\mathrm{out}},\nonumber\\
&b_{3,\mathrm{in}}=b_{2,\mathrm{out}}.
\end{align}

According to Ref. \cite{14.1}, it will be convenient to write the dynamics of the original model in
quadrature form as \eqref{dx2}
with
\begin{align}
&A=\nonumber\\&\left[\begin{array}{cccccc}-\frac{\kappa_1}{2}&\omega_1 &0&0&0&0  \\-\omega_1&-\frac{\kappa_1}{2}&0&0&0&0\\
-\sqrt{\kappa_1\kappa_2}&0&-\frac{\kappa_2}{2}&\omega_2&0&0
\\
0&\sqrt{\kappa_1\kappa_2}&-\omega_2&-\frac{\kappa_2}{2}&0&0\\
-\sqrt{\kappa_1\kappa_3}&0&-\sqrt{\kappa_2\kappa_3}&0&-\frac{\kappa_3}{2}&\omega_3\\
0&-\sqrt{\kappa_1\kappa_3}&0&-\sqrt{\kappa_2\kappa_3}&-\omega_3&-\frac{\kappa_3}{2}\end{array}\right],\nonumber
\end{align}
\begin{align}
B&=\left[\begin{array}{cc}-\sqrt{\kappa_1}&0\\0&-\sqrt{\kappa_1}\\-\sqrt{\kappa_2} &0\\0&-\sqrt{\kappa_2}\\-\sqrt{\kappa_3} &0\\0&-\sqrt{\kappa_3}
\end{array}\right],\nonumber\\
C&=\left[\begin{array}{cccccc}\sqrt{\kappa_1}&0&\sqrt{\kappa_2}&0&\sqrt{\kappa_3}&0\\0&\sqrt{\kappa_1}&0&\sqrt{\kappa_2}&0&\sqrt{\kappa_3}
\end{array}\right]\nonumber,\\
D&=\left[\begin{array}{cc}1&0\\0&1
\end{array}\right]\nonumber.
\end{align}

It is easily determined that the original model fulfills the criteria for physical realizability as outlined in \eqref{realr21}, \eqref{realr22} and \eqref{realr23}. Our aim is to simplify the original model into a fourth-order system according to $\mathscr{H}_2$ norm performance, which also satisfies the physical realizable conditions.

Based on Theorem 5, we have derived the reduced model
as follows
\begin{align}
A_r&=\left[\begin{array}{cccc}
 -0.24  & -5.54   & 0.74 &   1.37\\
    4.89&  -0.99  &  2.66  & -8.03\\
   -0.63  & -1.54 & -0.51 &   1.52\\
   -0.91  &  7.93  & -1.39  & -1.37\\
\end{array}\right]\nonumber,\\
B_r&=\left[\begin{array}{cc}
  0.35 & -0.36\\
    0.32  & -1.31\\
   -0.19  &  0.84\\
   -1.60  & -0.46\\
\end{array}\right]\nonumber,\\
C_r&=\left[\begin{array}{cccc}
   -0.35 &  -0.32  &  0.19  &  1.60\\
    0.37 &   1.31  & -0.84  &  0.46\\
\end{array}\right]\nonumber.
\end{align}
%

\begin{figure}[htbp]
\centerline{\includegraphics[width=9cm]{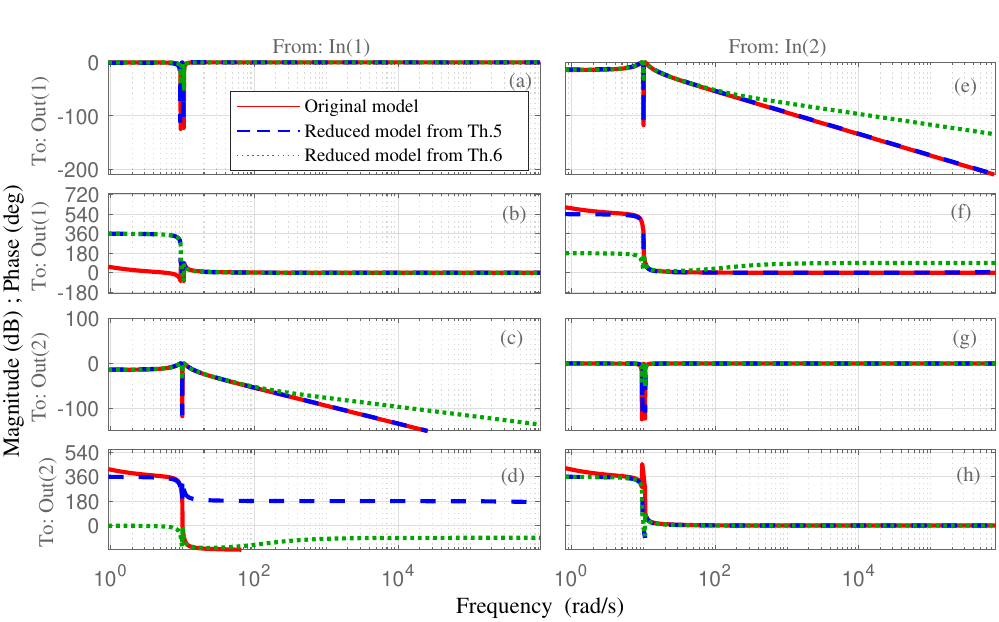}}
\caption{Comparison of frequency responses between the original and reduced models for a passive system}
\label{fig1}
\end{figure}

 By Theorem 5, we obtain $||\Xi_G-\Xi_{G_r}||^2_{2}=1.02$.
In addition, the Bode plots in  Fig. \ref{fig1} effectively demonstrate the comparison between the original model and the reduced model.
 Both magnitude and phase responses exhibit minimal deviations across the frequency range, suggesting that the reduced model successfully preserves the dynamic behavior of the system while potentially offering computational or analytical advantages.
 This consistency is especially evident in the magnitude plots Fig. \ref{fig1} (a), (c), (e) and (g), where the reduced model closely follows the original model's response, particularly in key frequency regions.
   Overall, the model reduction appears to have been performed with high accuracy, ensuring that the simplified model remains a reliable representation of the original system.

\section{Conclusion}
This paper has presented an optimal strategy for model reduction based on $\mathscr{H}_2$ performance, specifically designed for general linear quantum   systems.
The primary innovation lies in reformulating the model reduction problem as an optimization problem, followed by deriving necessary conditions for its solution.
A significant contribution is the transformation of challenging nonlinear constraints arising from physical realizability into  LMI constraints and solvable nonlinear equality constraints.
 Moreover, a scheme for model reduction for passive linear quantum systems is also presented.
The viability of our method is demonstrated through its application to active and passive linear quantum systems.
Most importantly, given that the inputs to linear quantum systems are often characterized by white noise, the $\mathscr{H}_2$ model reduction framework proposed in this paper is more appropriately aligned with practical scenarios.
Future work could venture into expanding these methodologies to include the optimization of $\mathscr{H}_\infty$ performance in model reduction, offering new avenues for enhancing system robustness and performance in quantum systems.

\vspace{12pt}

\end{document}